\documentclass[preprint,showpacs,preprintnumbers,amsmath,amssymb]{revtex4}
\usepackage{graphicx}
\usepackage{dcolumn}
\usepackage{bm}

\begin{document}

\title{Chirped dissipative solitons of the complex cubic-quintic nonlinear Ginzburg-Landau equation}

\author{V. L. Kalashnikov}
\affiliation{Institut f\"{u}r Photonik, TU Wien, Gusshausstr.
27/387, A-1040 Vienna, Austria} \email{kalashnikov@tuwien.ac.at}

\begin{abstract}
Approximate analytical chirped solitary pulse (chirped dissipative
soliton) solutions of the one-dimensional complex cubic-quintic
nonlinear Ginzburg-Landau equation are obtained. These solutions are
stable and highly-accurate under condition of domination of a normal
dispersion over a spectral dissipation. The parametric space of the
solitons is three-dimensional, that makes theirs to be easily
traceable within a whole range of the equation parameters. Scaling
properties of the chirped dissipative solitons are highly
interesting for applications in the field of high-energy ultrafast
laser physics.
\end{abstract}


\maketitle

\section{\label{intro}Introduction}

The complex nonlinear Ginzburg-Landau equation (CGLE) has so wide
scope of applications that the concept of ``the world of the
Ginzburg-Landau equation" \cite{kramer} is not exaggeration. The
CGLE demonstrates its effectiveness in quantum optics, modeling of
Bose-Einstein condensation, condensate-matter physics, study of
non-equilibrium phenomena, and nonlinear dynamics, quantum mechanics
of self-organizing dissipative systems, and quantum field theory. In
optics and laser physics, the CGLE provides an adequate description
of mode-locked oscillators and pulse propagation in fibers
\cite{a3,kartner1}.

The CGLE is multiparameter and not integrable in a general form. As
a result, an analysis of multitude of its solutions requires
extensive numerical simulations. The exact analytical solutions are
known only for a few of cases, when they represent the solitary
waves (dissipative solitons) and some algebraic relations on the
parameters of equation are imposed \cite{a3,a1}. As a rule, one
presumes some class of functional expressions to construct the
solution. As a result of such presupposition, the solutions outside
a given class are missed. In principle, the missed solutions can be
revealed on basis of the algebraic non-perturbative techniques
\cite{a1,c}, which, nevertheless, need a lot of computer algebra.
These challenges stimulate interest in the approximate methods of
integration of the CGLE.

The perturbative method has allowed obtaining the dissipative
soliton solutions of the reduced and complete cubic-quintic CGLE in
the limits of small and large dispersion and phase nonlinearity
\cite{m1,m2}. Another approximate approach is to reduce an
infinite-dimensional (in terms of degrees of freedom) problem to
finite-dimensional one on basis of, for instance, the method of
moments. This allows tracing an evolution of a finite set of the
trial solution parameters \cite{a2}.

However, some physically interesting sectors of the CGLE allow an
approximation without any functional constraints imposed on the
solution or/and the equation parameters. Moreover, dimensionality of
the parametric space corresponding to such a solution can be reduced
in comparison with the parametric space of the CGLE that makes the
solution under consideration to be easily traceable.

A physically important sector, which permits an approximate
analysis, is represented by the chirped solitary pulse solutions, or
the chirped dissipative solitons (CDSs) of the CGLE. The CDS exists
in both anomalous and normal dispersion ranges \cite{a2,h,a4}. It is
very important, that the CDS is energy scalable \cite{a2,a4,k1,a5}
and can be considered as a model of femtosecond laser pulses with
about of and over microjoule energies \cite{ap,m}. Energy
scalability of the CDS results from its stretching caused by a large
chirp. Hence, the CDS with large energy has a reduced peak power
that provides its stability \cite{k2,k3}. Simultaneously, a large
chirp leads to spectral extra-broadening so that the CDS becomes to
be compressible down to a few of tens of femtoseconds
\cite{ap,k1,k2}.

The mechanism of the CDS formation is a composite balance of phase
and dissipative effects \cite{a2}. The first effect is a balance of
phase contributions from the nonlinear phase distortion and the
time-dependent phase affected by a normal dispersion \cite{k4}. That
is possible if a soliton is chirped, but this effect alone does not
provide a soliton stability. The CDS stability can be provided by a
balance between the nonlinear gain and the spectral dissipation
\cite{h,k4,p}.

A large chirp of the CDS allows two main approximations: i) soliton
stretching admits the adiabatic approximation, ii) fast phase
variation allows applying the stationary phase method in Fourier
domain \cite{k2,k3,k5}. As a result, the CDS of the reduced
cubic-quintic one-dimensional CGLE \cite{k2,k5} and the generalized
one-dimensional CGLE (i.e. the CGLE with an imaginary saturable law
nonlinearity in terms of \cite{nonkerr}) \cite{k3} can be
represented analytically as the two-parametric solitary pulse
solution without any restrictions on its functional form as well as
on the equation parameters (certainly, within the scope of
approximations under consideration, see below).

Here the extension of this approximate technique to the complete
cubic-quintic one-dimensional CGLE (i.e. the CGLE with a complex
parabolic law nonlinearity \cite{nonkerr}) is presented. It is
shown, that the CDS is the three-parametric solution with five types
of the truncated spectral profiles: i) finger-, ii) parabolic-, and
iii) flat-top, as well as iv) concave and v) concave-convex ones.
The regions of existence and stability of the CDS are analyzed
systematically within a whole parametric range obeying the condition
of domination of a normal dispersion over a spectral dissipation.
The obtained results are validated on the basis of numerical
solution of the CGLE and compared with the existing results of
extensive numerical simulations of the CGLE.

\section{\label{soliton}CDS of the cubic-quintic CGLE}

Let the CGLE be written down in the following form \cite{k2,k5}:

\begin{widetext}
\begin{equation}\label{GL}
\frac{\partial } {{\partial z}}a\left( {z,t} \right) =  - \sigma
a\left( {z,t} \right) + \left( {\alpha  + i\beta }
\right)\frac{{\partial ^2 }} {{\partial t^2 }}a\left( {z,t} \right)
+ (\kappa  - i\gamma )\left| {a\left( {z,t} \right)} \right|^2
a\left( {z,t} \right) - \left( {\kappa \zeta  + i\chi }
\right)\left| {a\left( {z,t} \right)} \right|^4 a\left( {z,t}
\right).
\end{equation}
\end{widetext}

\noindent Here $z$ is the propagation (longitudinal) coordinate,
which can be the propagation distance in a fiber, or the cavity
round-trip in a laser oscillator, for instance; $t$ is the
``transverse" coordinate, which can be, for instance, the local time
for a propagating laser pulse \cite{a3}. The complex slowly-varying
field amplitude $a(z,t)$ is chosen so that $|a|^2$ has a dimension
of instant power. The first term on right-hand side of Eq.
(\ref{GL}) describes an action of net-loss with the parameter
$\sigma$. In the general case, this parameter is energy-dependent
(i.e. it depends on $ \int_{ - \infty }^\infty  {\left| {a\left(
{z,t'} \right)} \right|} ^2 dt' $, see \cite{k2}) and has to be
positive to provide the vacuum stability [ i.e. the subcritical
range of Eq. (\ref{GL}) is under consideration]. The second term
describes a spectral dissipation ($\alpha$ is the squared inverse
bandwidth of spectral filter) and a dispersion [$\beta$ is the
dispersion coefficient]. Positivity (negativity) of $\beta$
corresponds to normal (anomalous) dispersion. The third term results
from a contribution of cubic nonlinearity, which is a sum of
contributions from the nonlinear gain [or the self-amplitude
modulation (SAM) defined by the parameter $\kappa>$0] and from the
self-phase modulation (SPM) with the parameter $\gamma$. Only the
focusing SPM with $\gamma>$0 will be considered below. A
higher-order (quintic) nonlinearity defines the fourth term in Eq.
(\ref{GL}). Its real part describes the SAM saturation ($\zeta>$0
provides stability of a desired solution against collapse), while a
correction to the cubic SPM can be both enhancing ($\chi>$0) and
saturating ($\chi<$0).

To find the CDS solution, let's make the traveling wave reduction of
Eq.(\ref{GL}) by means of the ansatz

\begin{equation}\label{ansatz}
a\left( {z,t} \right) = \sqrt {P\left( t \right)} \exp \left[ {i\phi
\left( t \right) - iqz} \right],
\end{equation}

\noindent where $P(t)$ is the instant power, which defines a CDS
envelope; $\phi(t)$ is the phase, and $q$ is the phase shift due to
a slip of the carrier phase with respect to an envelope
\cite{kartner1}.

Below we shall consider only the sector of CGLE, where a normal
dispersion prevails over a spectral dissipation, that is $\beta \gg
\alpha >$0 \cite{m1}. This assumption is well-grounded for both
broadband solid-state \cite{k1,k2,k3} and fiber \cite{k3,w1,w2}
laser oscillators operating in the all-normal dispersion (ANDi)
regime. But the numerical analysis demonstrates \cite{k3}, that even
the case of $\beta \geq \alpha >$0 (e.g., a thin-disk solid-state
oscillator \cite{m}) can be described adequately in the framework of
the analytical approach under consideration.

The adiabatic approximation $T \gg \sqrt{\beta}$ allows obtaining
from Eqs. (\ref{GL},\ref{ansatz})

\begin{eqnarray} \label{reduced}
\begin{gathered}
  \beta \Omega ^2  = q - \gamma P - \chi P^2 , \hfill \\
  \beta \left( {\frac{\Omega }
{{P\left( t \right)}}\frac{{dP}} {{dt}} + \frac{{d\Omega }}
{{dt}}} \right) = \kappa P\left( {1 - \zeta P} \right) - \sigma  - \alpha \Omega ^2 , \hfill \\
\end{gathered}
\end{eqnarray}

\noindent where $\Omega \equiv {{d\phi \left( t \right)}
\mathord{\left/
 {\vphantom {{d\phi \left( t \right)} {dt}}} \right.
 \kern-\nulldelimiterspace} {dt}}$ is the instant frequency.

Since the first equation in (\ref{reduced}) is quadratic in $P$,
there are two branches of solution. However, it is reasonable to
confine oneself to the branch, which has the limit $\chi
\rightarrow$0 (this limit has been considered in \cite{k2,k5}).
Then, one has

\begin{equation}\label{power}
P =  - \frac{1} {{2\chi }}\left( {\gamma  - \sqrt {\gamma ^2  +
4q\chi  - 4\chi \beta \Omega ^2 } } \right).
\end{equation}

Since $P\geq$0 by definition, there is the maximum frequency
deviation $\Delta$ from the carrier frequency: $ \Delta ^2  = {q
\mathord{\left/
 {\vphantom {q \beta }} \right.
 \kern-\nulldelimiterspace} \beta }
$. Thus, the second equation in (\ref{reduced}) and Eq.
(\ref{power}) lead, after some algebra \cite{html}, to

\begin{eqnarray} \label{omega}
\begin{gathered}
  \frac{{d\Omega }}
{{dt}} = \frac{{\sigma  + \alpha \Omega ^2  - \frac{{\kappa \left(
{A - \gamma } \right)}} {{4\chi ^2 }}\left( {2\chi  + \zeta \gamma
- \zeta A} \right)}}
{{\beta \left[ {4\chi \beta \Omega ^2  - \left( {A - \gamma } \right)A} \right]}}\left( {A - \gamma } \right)A, \hfill \\
  A = \sqrt {\gamma ^2  + 4\beta \chi \left( {\Delta ^2  - \Omega ^2 } \right)} . \hfill \\
\end{gathered}
\end{eqnarray}

The singularity points of Eq. (\ref{omega}) impose the restrictions
on the $\Delta$ value \cite{html}

\begin{widetext}
\begin{equation}\label{delta}
\Delta ^2  = \frac{\gamma } {{16\zeta \beta \left( {\frac{c} {b} +
1} \right)}}\left[ {\frac{{2\left( {3 + \frac{c} {b} + \frac{4} {b}}
\right)\left( {2 + \frac{c} {2} + \frac{{3b}} {2} \pm \sqrt {\left(
{c - 2} \right)^2  - 16 a \left( {1 + \frac{c} {b}} \right)} }
\right)}} {{1 + \frac{c} {b}}} - 12 - 3c - 9b - \frac{{32a }} {{
b}}} \right],
\end{equation}
\end{widetext}

\noindent where three control parameters are $ a \equiv {{\sigma
\zeta } \mathord{\left/
 {\vphantom {{\sigma \zeta } \kappa }} \right.
 \kern-\nulldelimiterspace} \kappa }
$, $ b \equiv {{\zeta \gamma } \mathord{\left/
 {\vphantom {{\zeta \gamma } \chi }} \right.
 \kern-\nulldelimiterspace} \chi }
$, and
 $ c \equiv {{\alpha
\gamma } \mathord{\left/
 {\vphantom {{\alpha \gamma } {\beta \kappa }}} \right.
 \kern-\nulldelimiterspace} {\beta \kappa }}
$. These three parameters define the parametric dimensionality of
the CDS. Eqs. (\ref{power},\ref{delta}) allow obtaining the CDS peak
power ($\Omega$ has to be equal to 0 in Eq. (\ref{power}) for this
aim).

It is convenient to use the following normalizations: $ t' = t\left(
{{\kappa  \mathord{\left/
 {\vphantom {\kappa  \zeta }} \right.
 \kern-\nulldelimiterspace} \zeta }} \right)\sqrt {{\kappa  \mathord{\left/
 {\vphantom {\kappa  {\alpha \zeta }}} \right.
 \kern-\nulldelimiterspace} {\alpha \zeta }}}
$, $ \Delta '^2  = {{\Delta ^2 \alpha \zeta } \mathord{\left/
 {\vphantom {{\Delta ^2 \alpha \zeta } \kappa }} \right.
 \kern-\nulldelimiterspace} \kappa }
$, $ \Omega '^2  = {{\Omega ^2 \alpha \zeta } \mathord{\left/
 {\vphantom {{\Omega ^2 \alpha \zeta } \kappa }} \right.
 \kern-\nulldelimiterspace} \kappa }
$, $ P' = \zeta P$. For the dimensionless energy, one has $ E' =
E\left( {{\kappa  \mathord{\left/
 {\vphantom {\kappa  \gamma }} \right.
 \kern-\nulldelimiterspace} \gamma }} \right)\sqrt {{{\kappa \zeta } \mathord{\left/
 {\vphantom {{\kappa \zeta } \alpha }} \right.
 \kern-\nulldelimiterspace} \alpha }}
$. Hereinafter, these normalization will be implied and the primes
will be omitted.

The expressions for the dimensionless quantities are

\begin{widetext}
\begin{eqnarray} \label{system}
\begin{gathered}
  A = \sqrt {1 + \frac{{4\left( {\Delta ^2  - \Omega ^2 } \right)}}
{{cb}}} , \hfill \\
  P = \frac{b}
{2}\left( {\sqrt {1 + \frac{{4\left( {\Delta ^2  - \Omega ^2 }
\right)}}
{{cb}}}  - 1} \right), \hfill \\
  \frac{{d\Omega }}
{{dt}} = \frac{{c\left[ {a + \Omega ^2  + \frac{{b^2 }} {4}\left( {1
- A} \right)\left( {\frac{2} {b} + 1 - A} \right)} \right]A\left( {A
- 1} \right)}} {{\frac{{4\Omega ^2 }}
{{cb}} + A\left( {1 - A} \right)}}, \hfill \\
  \Delta ^2  = \frac{c}
{{16\left( {\frac{c} {b} + 1} \right)}}\left[ {\frac{{2\left(
{\frac{c} {b} + 3 + \frac{4} {b}} \right)\left( {\frac{c} {2} +
\frac{{3b}} {2} + 2 \pm \sqrt {\left( {c - 2} \right)^2  - 16a\left(
{\frac{c} {b} + 1} \right)} } \right)}} {{\frac{c} {b} + 1}} - 3c -
9b - \frac{{32a}}
{b} - 12} \right]. \hfill \\
\end{gathered}
\end{eqnarray}
\end{widetext}

Since the phase $\phi(t)$ of the CDS can be treated as a rapidly
varying function of time in the limit of $\kappa \ll \gamma$, one
may apply the method of stationary phase to the Fourier image of
$a(t)$ \cite{k5}. As a result, the expression for the CDS spectral
profile is

\begin{widetext}
\begin{equation}
p\left( \omega \right) \equiv |e\left( \omega \right) |^{2} \approx
\frac{\pi \left( A-1\right) \left( \left( A-1\right) cb+4\left(
2\omega ^{2}-\Delta ^{2}\right) \right) H\left( \Delta ^{2}-\omega
^{2}\right) }{cA\left( \left( A-1\right) \left( c\left(
a+b+b^{2}+\omega ^{2}\right) +b\left( \Delta ^{2}-\omega ^{2}\right)
\right) -2\left( b+1\right) \left( \Delta ^{2}-\omega ^{2}\right)
\right) }. \label{spectrum}
\end{equation}
\end{widetext}

\noindent where $e\left( \omega \right) \equiv
\int dt\sqrt{P\left( t\right) }\exp \left[ i\phi \left( t\right) -i\omega t%
\right] $, $H\left( x\right) $ is the Heaviside's function and one has
to replace $\Omega$ by $\omega$ in $A$ given by Eq. (\ref{system}).

The CDS energy can be obtained from Eq. (\ref{spectrum}) by
integration: $ E = \int_{ - \Delta }^\Delta  {\frac{{d\omega }}
{{2\pi }}p\left( \omega  \right)}$. This value can be related to the
energy $E^*$ of a solution of the linearized version of Eq.
(\ref{GL}) through the saturated net-loss parameter $\sigma$:
$\sigma \approx \delta (E/E^*-1)$, where $ \delta  \equiv \left.
{{{d\sigma } \mathord{\left/
 {\vphantom {{d\sigma } {dE}}} \right.
 \kern-\nulldelimiterspace} {dE}}} \right|_{E = E^* }
$ \cite{k2}. Such a relation can be usable, for instance, to define
the CDS parameters from those of a laser oscillator \cite{k2,k3}.

Thus, the approximate technique under consideration allows
representing the CDS parameters, its spectral and temporal profiles
as well as energy from a few of algebraic expressions
(\ref{system},\ref{spectrum}), single first-order ordinary
differential equation (\ref{system}) and numerical integration of
(\ref{spectrum}). Since the CDS is three-parametric, such an
approximation allows easily tracing the soliton characteristics
within a broad range of the CGLE parameters. It is important, that
the absolute values of the CGLE parameters are not relevant in
contrast to their relations presented by the parameters $a$, $b$ and
$c$. This allows a unified viewpoint at the diverse systems obeying
the CGLE \cite{k3,k6}.

\section{\label{md}Master diagram and regions of the CDS existence}

Eq. (\ref{system}) demonstrates that there exist two branches of the
CDS corresponding to two signs before square root in the expression
for $\Delta$. As will be shown below, such a division into two
branches is physically meaningful. In accordance with the sign in
Eq. (\ref{system}), we shall denote these branches as the
``positive" ($+$) and ``negative'' ($-$) ones. One has note, that
only the $-$ branch has a limit for $\zeta, \chi \rightarrow 0$ (the
``Schr\"{o}dinger limit'') \cite{k5}.

\subsection{Positive branch of the CDS}

\begin{figure}
\includegraphics[width=8.5cm]{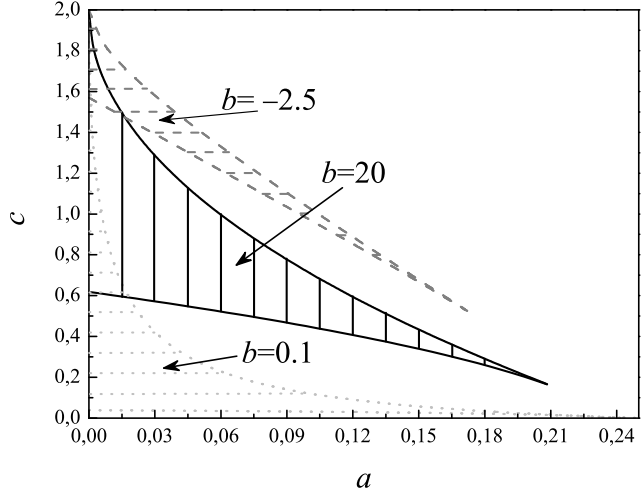}
\caption{\label{fig1} Regions (hatched) of the $+$ branch
existence. $b=$20 (black solid curves, vertical hatching), 0.1
(light gray dotted curves, horizontal hatching), and $-$2.5 (gray dashed
curves, horizontal hatching).}
\end{figure}

Regions of the $+$ branch existence are shown in Fig. \ref{fig1} on
the plane ($a$--$c$) for the different $b$. Zero value of $a$
corresponds to a marginally stable CDS. The existence regions are
maximally broad in this case. The restrictions on the parameters are
$0<c<2$, and $b>$0 (the SPM is unsaturable) or $b < -c/3 - 4/3$ (the
SPM is saturable).  One has to resemble, that the decrease of $\vert
b \vert$ means a growth of contribution of the quintic SPM. The
physical meaning of maximum $c$ is that there is a minimum
dispersion or a maximum spectral dissipation, which provides the CDS
existence.

A new view on the CDS results from a consideration of chirp at the
soliton center (i.e. at $t=$0):

\begin{eqnarray}\label{chirp}
\begin{gathered}
\left. \psi  \right|_{t = 0}  \equiv \left. {\frac{{d\Omega }}
{{dt}}} \right|_{\Omega  = 0}  =  - a - \frac{{b^2 }} {4}\left( {1 -
\left. A \right|_{\Omega  = 0} } \right) \\
\times\left( {1 + \frac{2} {b} - \left. A \right|_{\Omega  = 0} }
\right).
\end{gathered}
\end{eqnarray}

\noindent Here the chirp is normalized to $ {\kappa \mathord{\left/
 {\vphantom {\kappa  {\zeta \beta }}} \right.
 \kern-\nulldelimiterspace} {\zeta \beta }}
$. The analysis demonstrates that the chirp becomes negative, when
the $c$ parameter reaches some minimum value (the lower borders of
the hatched regions in Fig. \ref{fig1}). Such a chirp corresponds to
a spike on constant background with $ \mathop {\lim }\limits_{t
\to  \pm \infty } \Omega < \pm \infty$, $ \mathop {\lim }\limits_{t
\to  \pm \infty } P = const > 0 $. These solutions will be not considered
hereafter and the chirp positivity will be admitted as the additional criterion of the CDS existence.
This criterion agrees with the analytical results of \cite{a4} in the limit of
$\alpha/\beta \ll$1 and $\kappa/\gamma \ll$1. The appropriate zero
asymptotic $ \mathop {\lim }\limits_{t
\to  \pm \infty } P = 0 $ of the solutions analyzed in \cite{a4} exists only if $0 < \psi
\approx 3\gamma/(1+c)\kappa < \beta/\alpha$. Here $\psi$ is defined
as the parameter in the phase profile ansatz $\phi(t)=\psi
\ln{\sqrt{P(t)}}$, which is used in \cite{a4}.

As a result, there is some minimum $c$ (i.e. maximum normal
dispersion or minimum spectral dissipation) for the $+$ branch (Fig.
\ref{fig1}). This minimum $c$ tends to zero, when the positive $b$
decreases (Fig. \ref{fig1}).

For $b<$0, the CDS existence range squeezes, when $b$ approaches
$-2$ (Fig. \ref{fig1}). If $b>-$2, the positively chirped CDS has a
parabolic temporal profile and $ \mathop {\lim }\limits_{t \to \pm
\tau } \Omega  = \pm \infty$, $ \mathop {\lim }\limits_{t \to \pm
\tau } P = 0$ ($\tau$ is some finite interval of local time). We
will not consider such an ``inverted'' CDS hereafter.

The $a$ growth, if it results from the $\sigma$ increase,
enhances the soliton stability against a vacuum destabilization.
However, the existence regions shrink along the $c$ parametric
coordinate with such a growth (Fig. \ref{fig1}).

\subsection{Negative branch of the CDS}

\begin{figure}
\includegraphics[width=8.5cm]{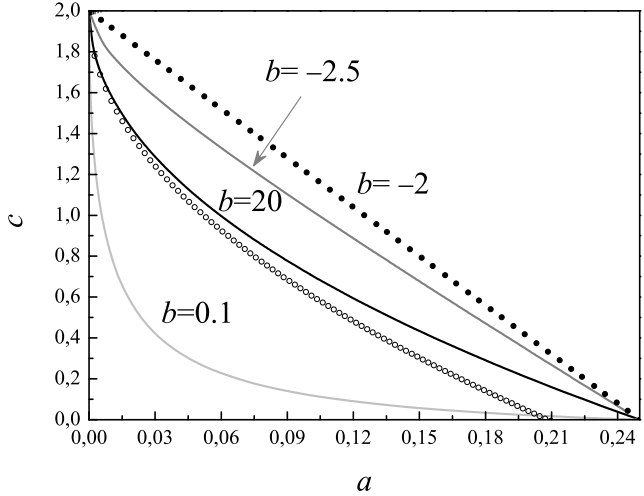}
\caption{\label{fig2} Borders of regions of the $-$ branch existence
(black solid curve for $b=$20, light gray solid curve for $b=$0.1,
gray solid curve for $b=-$2.5, points for $b=-$2). Open circles
correspond to the region border from \cite{m1}. The existence
regions lie below the corresponding borders.}
\end{figure}

Regions of the $-$ branch existence are shown in Fig. \ref{fig2} on
the plane ($a$--$c$) for the different $b$. The CDS exists within
the interval 0$<c<$2, which squeezes with $a$. Since this branch has
a Schr\"{o}dinger limit, such a squeezing can be obtained on the
basis of the perturbative method \cite{m1}. Then, the existence
region for $|b|\gg 1$ is $ c \leqslant 2 - 4\sqrt {{{6a}
\mathord{\left/
 {\vphantom {{6a} 5}} \right.
 \kern-\nulldelimiterspace} 5}}
$ (open circles in Fig. \ref{fig2}) \cite{m1}. One can see, that the
approximation of \cite{m1} is quite accurate, when $a \ll 1$ (i.e.
in the low-energy limit). The limiting $a$ is defined by the hard
excitation condition $ a \leqslant {1 \mathord{\left/
 {\vphantom {1 4}} \right.
 \kern-\nulldelimiterspace} 4}$ \cite{m1}. The existence region shrinks with a growing positive quintic SPM (i.e., when
$b>$0 tends to zero) and stretches with a negative quintic SPM
verging towards $b=-$2.

There are no negative chirp solutions for this branch. There exist
the positive chirp solutions for $c>$2 and $b<-$2, but they are the
spikes on background.

One can see that the upper (in the $c$ direction) borders of the regions
coincide for the positive and negative branches. This means that the
branches coexist within the regions of their existence in the ($a$,
$b$, $c$)-parametric space.

\subsection{Master diagram}

Representation of the existence regions in the form of Figs.
\ref{fig1},\ref{fig2} is awkward in some way, because
the $a$ parameter can be energy-dependent. As a result, the branches do not coexist as they differ in energy. It
is more convenient to use a representation on the plane ($E$--$c$)
for the different $b$. Such a representation will be called
the \emph{master diagram}. The $E$ value can be easily related to the
experimentally controllable parameter $E^*$ (see Section
\ref{soliton}).

\begin{figure}
\includegraphics[width=8.5cm]{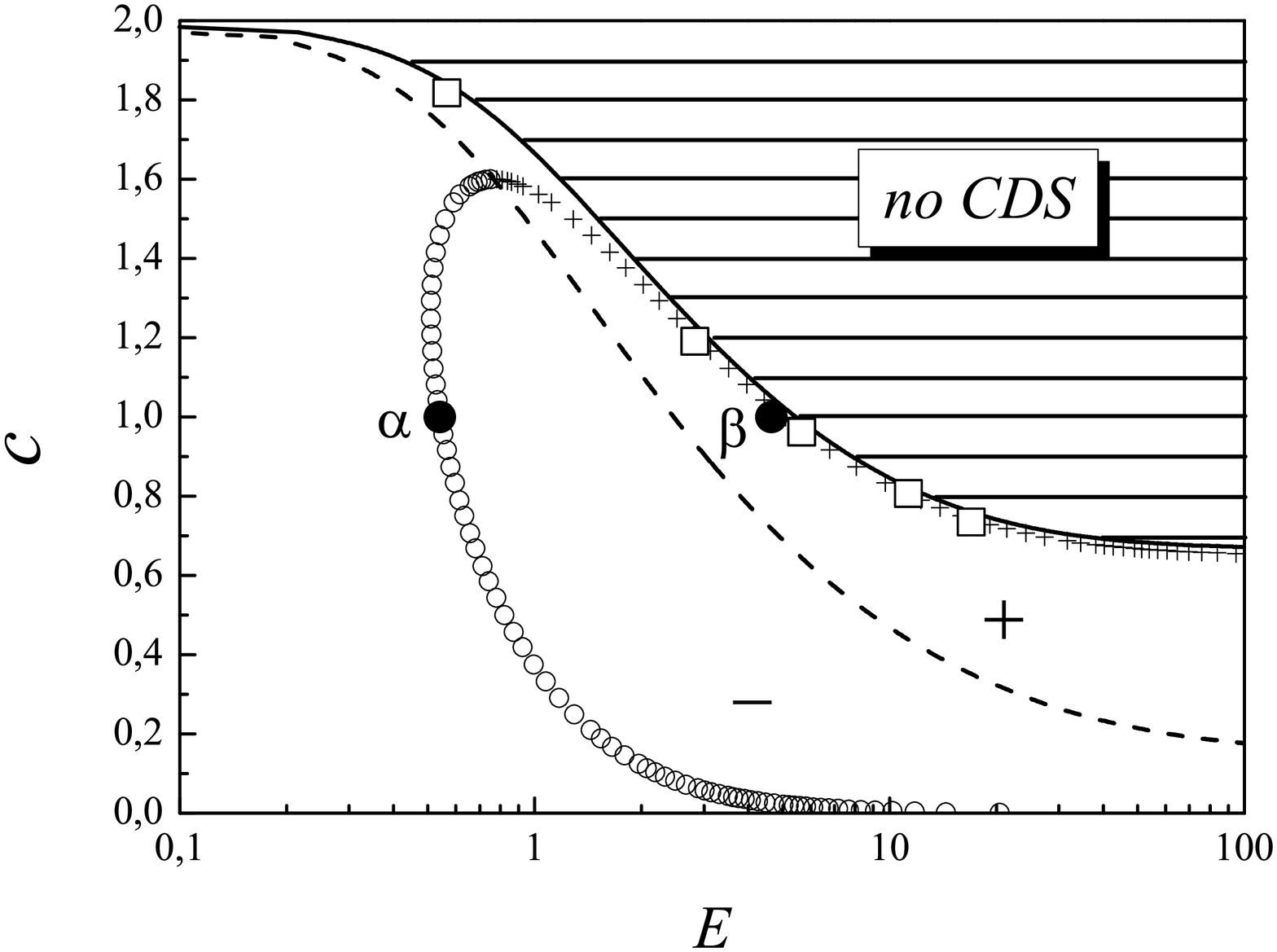}
\caption{\label{fig3} The master diagram for $b \gg$1. There exists
no CDS within the hatched region. Solid curve corresponds to $a=$0.
Dashed curve divides the regions, where the $+$ and $-$ branches
exist. Crosses (circles) correspond to the $+$ ($-$) branch for
$a=$0.01. The points $\alpha$ and $\beta$ correspond to the
parameters of the numerical solutions presented in Figs.
\ref{fig8},\ref{fig9} by open circles. The open squares indicate the
numerically obtained stability border ($\kappa=$0.04$\gamma$,
$\zeta=$0.5$\gamma$).}
\end{figure}

The master diagram for the CDS is shown in Fig. \ref{fig3} for the
case of vanishing quintic SPM ($b \gg$1 \cite{k2,k6}). The solid
curve shows the border of the CDS existence ($a=$0). Above this
border, the vacuum of Eq. (\ref{GL}) is unstable (hatched region).
The dashed curve divides the existence regions for the $+$ and $-$
branches (the branches merge along this curve). Crosses (circles)
represent the curve along which there exists the $+$ ($-$) branch
for some fixed value of $a$ (so-called the \textit{isogain curve}).

The master diagram is interrelated with the existence regions in
Figs. \ref{fig1},\ref{fig2}. The point of intersection of isogain
with the dashed curve defines the maximum value of $c$ in Figs.
\ref{fig1},\ref{fig2} for the corresponding $a$. Since the $+$
branch isogain has a nonzero asymptotic for $E\rightarrow \infty$,
there is the nonzero minimum $c$, which confines the $+$ branch
region
 for a fixed $a$ in Fig. \ref{fig1}. The $-$ branch has a zero asymptotic
for $E\rightarrow \infty$. Hence, the $-$ branch extends down to
$c=$0.

The master diagram reveals four significant differences between the
branches. The first one is that the $-$ branch has lower energy than
the $+$ branch for a fixed $c$. The second difference is that the
$+$ branch isogain has nonzero asymptotic for $E\rightarrow \infty$.
In this sense, the $+$ branch is energy scalable, that is its energy
growth does not require a substantial change of $c$. The $-$ branch
is not energy scalable, that is its energy growth needs a
substantial decrease of $c$ (e.g., owing to a dispersion growth)
\cite{k3}. The third difference is that the $+$ branch verges on
$\sigma=$0 within a whole range of $E$. The fourth difference is
that the $-$ branch has a Schr\"{o}dinger limit $\zeta, \chi
\rightarrow 0$.

\begin{figure}
\includegraphics[width=8.5cm]{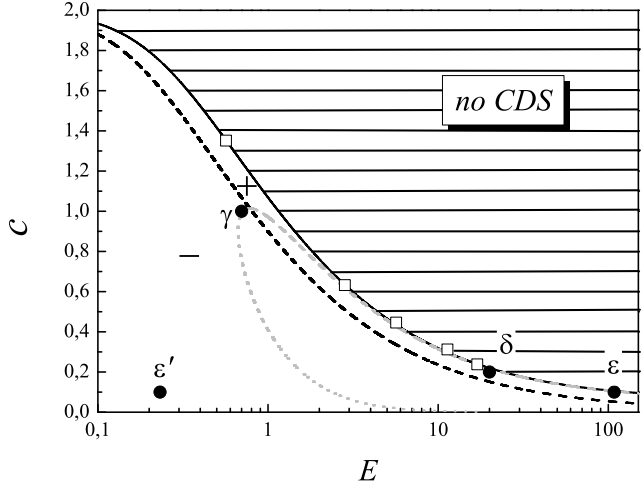}
\caption{\label{fig4} The master diagram for $b =$0.2. There exists
no CDS within the hatched region. Black solid curve corresponds to
$a=$0. Black dashed curve divides the regions, where the $+$ and $-$
branches exist. Gray dashed (dotted) curve corresponds to the $+$
($-$) branch for $a=$0.01. The point $\gamma$ corresponds to the
parameters of the numerical solution presented in Fig. \ref{fig9} by
open squares. The open squares indicate the numerically obtained
stability border ($\kappa=$0.04$\gamma$, $\zeta=$0.5$\gamma$). The
points $\delta$ and $\varepsilon$ correspond to the analytical
spectra presented in Fig. \ref{fig12} ($\kappa=$0.8$\gamma$ and
1.5$\gamma$, respectively; $\beta/\alpha=$6.25,
$\zeta=$0.002$\gamma$, $\sigma=$0.1 \cite{a5}). The point
$\varepsilon '$ is the $-$ branch counterpart of $\varepsilon$.}
\end{figure}

Growth of the positive quintic SPM (i.e. $b \rightarrow 0$) narrows
the existence region (Fig. \ref{fig4}). This means that smaller $c$
is required to provide the CDS existence for some $E$. That is,
since the positive quintic SPM means an enhancement of the phase
nonlinearity, a SPM enhancement has to be compensated, for instance,
by a dispersion increase ($c\propto 1/\beta$). One can see from Fig.
\ref{fig4}, that the $+$ branch region
 narrows substantially with
$b\rightarrow$0 ($b>$0) within a whole range of $E$.

\begin{figure}
\includegraphics[width=8.5cm]{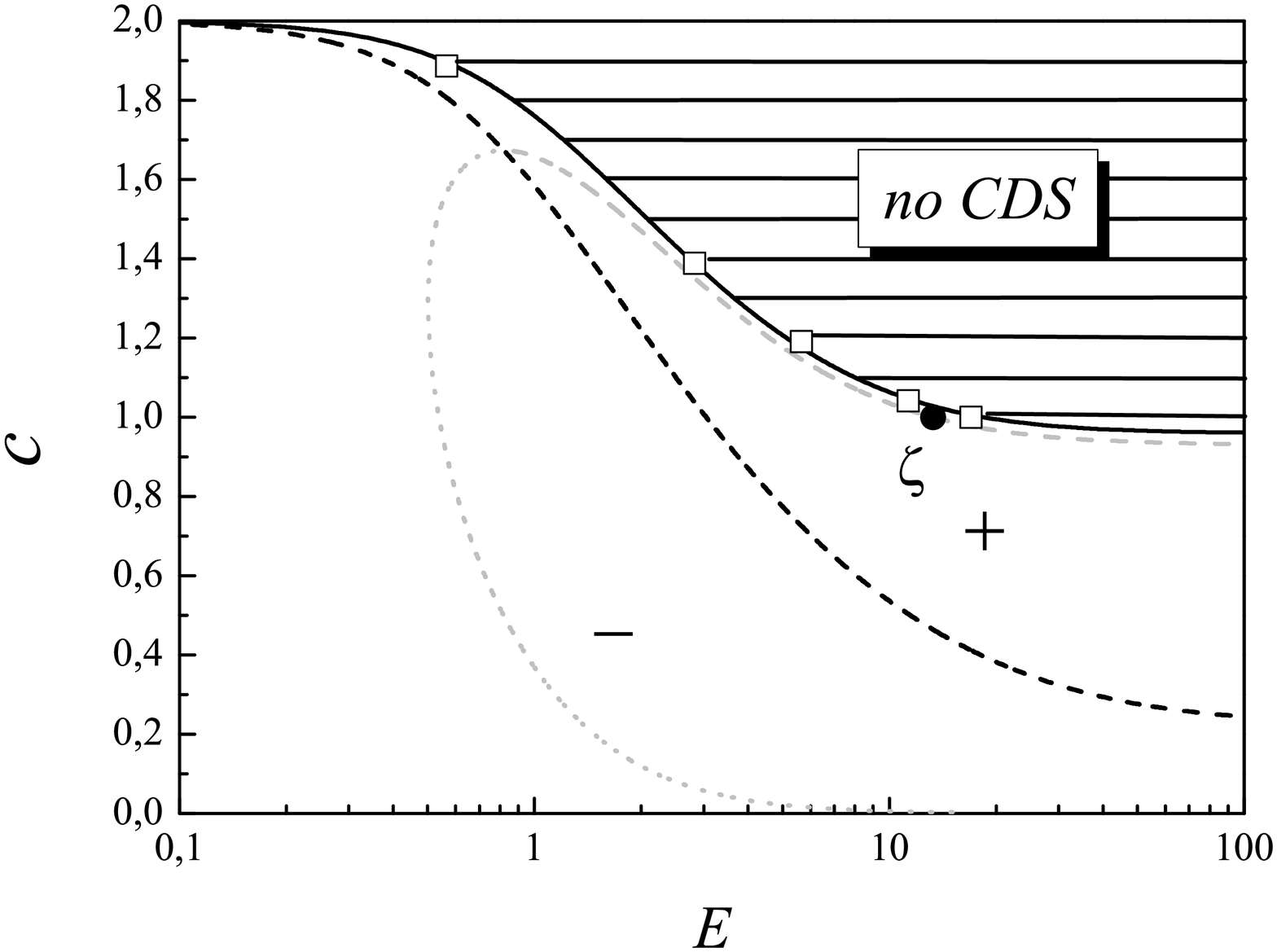}
\caption{\label{fig5} The master diagram for $b =$-5. There exists
no CDS within the hatched region. Black solid curve corresponds to
$a=$0. Black dashed curve divides the regions, where the $+$ and $-$
branches exist. Gray dashed (dotted) curve corresponds to the $+$
($-$) branch for $a=$0.01. The point $\zeta$ corresponds to the
parameters of the numerical solution presented in Fig. \ref{fig8} by
open squares. The open squares indicate the numerically obtained
stability border ($\kappa=$0.04$\gamma$, $\zeta=$0.5$\gamma$).}
\end{figure}

The situation is opposite, when the quintic SPM is negative. The
existence range widens and a larger $c$ (i.e. smaller dispersion)
provides the CDS existence for some $E$. The $+$ branch region
widens, as well. However, it is important to remember, that the
range of $c$, where the CDS with a fixed $a$ exists, is defined by
the difference between i) the point of intersection of the isogain
with the boundary between the $+$ and $-$ branches and ii) the
isogain asymptotic for $E\rightarrow \infty$. As a result, the range
of $c$, where some isogain exists, can be narrow in spite of the
fact that a whole range of the $+$ branch widens (Figs.
\ref{fig1},\ref{fig5}). The reversed situation, when a whole
existence range is narrow, but the range of $c$ for some isogain is
broad, is possible for $b>$0 (Figs. \ref{fig1},\ref{fig4}).

\section{\label{profile}CDS profile, spectrum and parameters}

\begin{figure}
\includegraphics[width=8.5cm]{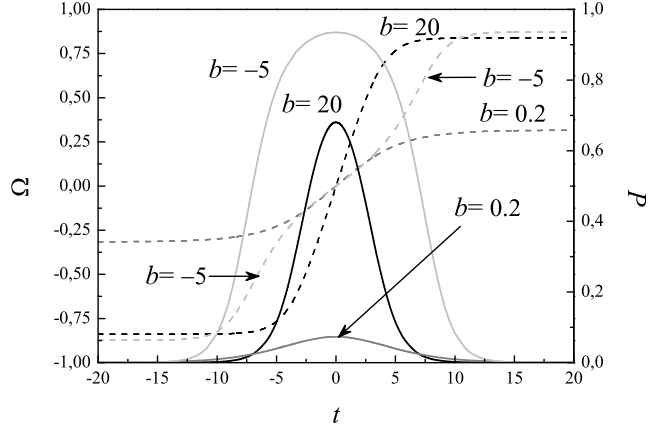}
\caption{\label{fig6} The $+$ branch CDS profiles (solid curves) and
frequency deviations (dashed curves) for the different $b$. $c=$1,
$a=$0.01.}
\end{figure}

Fig. \ref{fig6} shows the frequency deviations and the CDS profiles
relating to the $+$ branch  [see Eq. (\ref{system})] for the
different $b$. One can see, that the decrease of positive $b$
reduces a soliton energy (black solid vs. gray curves in Fig.
\ref{fig6}) for the fixed $c$ and $a$. That agrees with Figs.
\ref{fig3},\ref{fig4}, where the isogain shifts towards smaller
energies for a fixed $c$, when the positive $b$ tends to zero. Since
a power decreases, a chirp ($d\Omega/dt$) decreases, too (black vs.
gray dashed curves in Fig. \ref{fig6}).

In the case of $b<$0, the dependence of $\Omega$ on $t$ becomes
``loitering" (light gray dashed curve in Fig. \ref{fig6}). As a
consequence, the CDS profile becomes flat-top (light gray solid
curve in Fig. \ref{fig6}). The energy increases for a fixed $a$ in
agreement with a shift of the isogain towards larger energies in
Fig. \ref{fig5}.

\begin{figure}
\includegraphics[width=8.5cm]{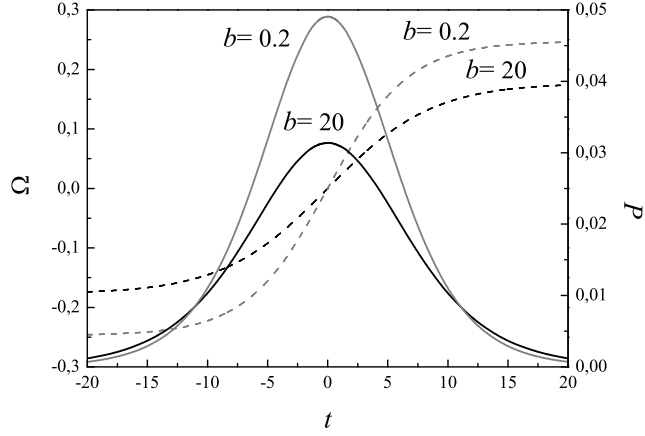}
\caption{\label{fig7} The $-$ branch CDS profiles (solid curves) and
frequency deviations (dashed curves) for the different $b$. $c=$1,
$a=$0.01.}
\end{figure}

The frequency deviations and the CDS profiles for the $-$ branch are
shown in Fig. \ref{fig7}. Out of the boundary between the branches,
the CDS relating to the $-$ branch has lower energy and power than
its $+$ counterpart. Correspondingly, a chirp is lower, as well.
Growth of the positive quintic SPM (i.e., $b \rightarrow$0)
increases the soliton energy, power and chirp (black vs. gray curves
in Fig. \ref{fig7}).

Growth of the negative quintic SPM decreases the CDS energy, power
and chirp. However, such a decrease is comparatively small and, therefore, it is not
shown in Fig. \ref{fig7}.

\begin{figure}
\includegraphics[width=8.5cm]{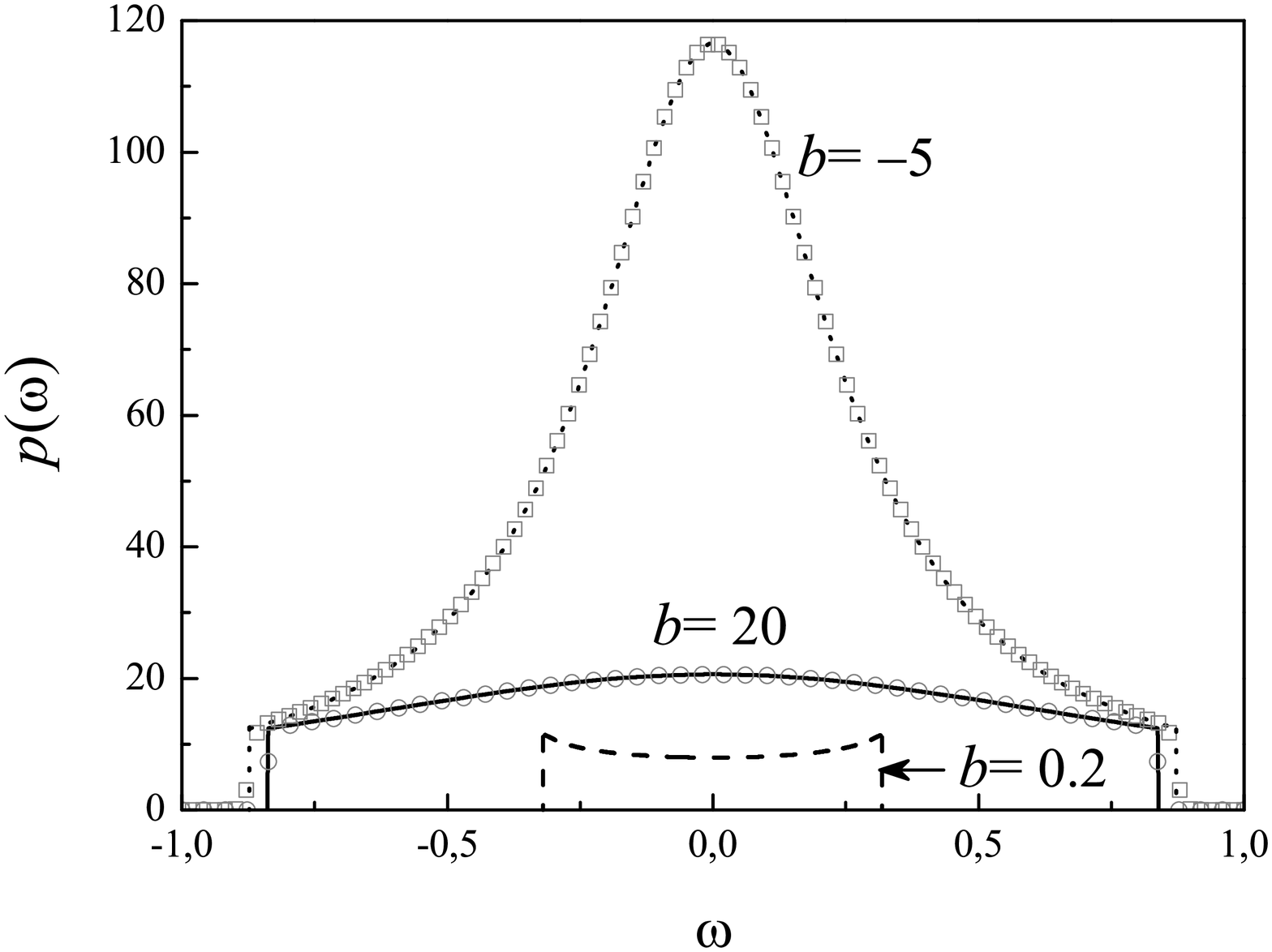}
\caption{\label{fig8} The $+$ branch CDS spectra for the different
$b$. $c=$1, $a=$0.01. Solid, dotted and dashed curves correspond to
analytical spectra. Gray circles correspond to the numerical
spectrum at the point $\beta$ in Fig. \ref{fig3} ($\zeta=0.5\gamma$,
$\beta/\alpha=$25, $\kappa=0.04\gamma$, $E=$820 $\kappa \sqrt{\kappa
\zeta}/\gamma^2$). Gray squares correspond to the numerical spectrum
at the point $\zeta$ in Fig. \ref{fig5} ($\zeta=0.5\gamma$,
$\beta/\alpha=$25, $\kappa=0.04\gamma$, $E=$2350 $\kappa
\sqrt{\kappa \zeta}/\gamma^2$).}
\end{figure}

\begin{figure}
\includegraphics[width=8.5cm]{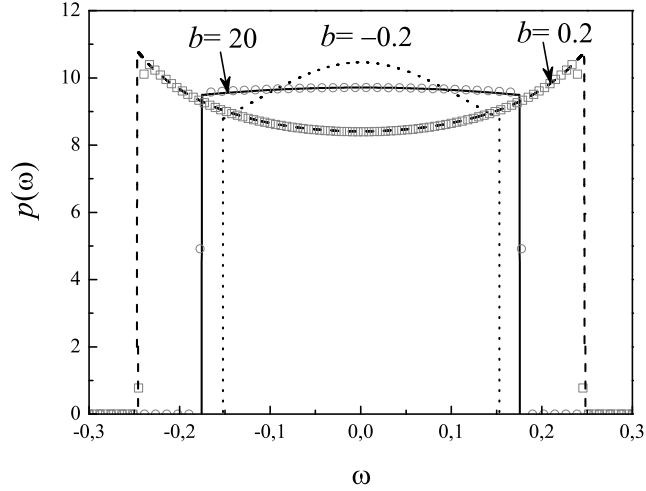}
\caption{\label{fig9} The $-$ branch CDS spectra for the different
$b$. $c=$1, $a=$0.01. Solid, dotted and dashed curves correspond to
the analytical spectra. Gray circles correspond to the numerical
spectra at the point $\alpha$ in Fig. \ref{fig3} (two coinciding
numerical profiles are defined by $\zeta=0.1\gamma$;
$\beta/\alpha=$30 and 40; $\kappa=0.033\gamma$ and $0.025\gamma$;
$E=$280 $\kappa \sqrt{\kappa \zeta}/\gamma^2$ and $E=$430 $\kappa
\sqrt{\kappa \zeta}/\gamma^2$, respectively). Gray squares
correspond to the numerical spectrum at the point $\gamma$ in Fig.
\ref{fig4} ($\zeta=0.1\gamma$, $\beta/\alpha=$42,
$\kappa=0.024\gamma$, $E=$600 $\kappa \sqrt{\kappa
\zeta}/\gamma^2$).}
\end{figure}

The CDS spectra are presented in Figs. \ref{fig8},\ref{fig9}. As
has been shown in Section \ref{soliton}, the spectra are
truncated at some frequency $\pm \Delta$. There are the following
types of spectral profiles: i) flat-top (solid curve in Fig.
\ref{fig9}), ii) convex (solid curve in Fig. \ref{fig8} and dotted
curve in Fig. \ref{fig9}), iii) finger-like (dotted curve in Fig.
\ref{fig8}) and iv) concave (dashed curves in Figs.
\ref{fig8},\ref{fig9}). All these types are widely presented in
laser experiments and numerical simulations
\cite{k1,a5,ap,k2,k3,w1}.

One can see, that, as a rule, the CDS spectra relating to the $+$
branch (Fig. \ref{fig8}) are broader than those relating to the $-$
branch (Fig. \ref{fig9}). The cause of this difference is a smaller
chirp for the $-$ branch CDS. The spectrum narrows (widens) with an
approach of positive $b$ to zero for the $+$ ($-$) branch in
accordance with a decrease (increase) of the CDS chirp. When the
positive quintic SPM increases ($b\rightarrow$0), concave spectra
appear. In contrast to the model of \cite{w2}, the source of such
spectra is not the self-amplifying SAM [i.e., the negative $\zeta$
in Eq. (\ref{GL})]
 but solely the
positive quintic SPM \cite{k6}. As a result, the concave spectrum
solution of Section 2 is stable against collapse.

It is important to note, that a verging of $b$ towards zero for the
$+$ branch as well as a transition to the $-$ branch reduce chirp.
This can violate a validity of the method of stationary phase (see
Section \ref{soliton}). As a result, the spectrum edges become
smooth (see Section \ref{compar}).

\begin{figure}
\includegraphics[width=8.5cm]{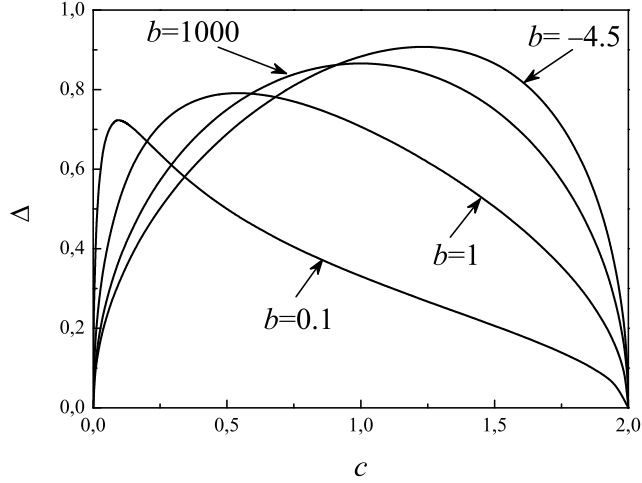}
\caption{\label{fig10} The CDS spectral half-widths for the $+$
branch and the different $b$. $a=$0.}
\end{figure}

As was mentioned earlier, the $+$ branch does not vanish along the
curve of $\sigma=$0. This curve corresponds to marginal stability
against a vacuum excitation and the CDS has a broadest spectrum
here. The dependence of half-width of such a spectrum on the $c$
parameter for the different $b$ are shown in Fig. \ref{fig10}. In
the absence of the quintic SPM ($b\gg$1), the dependence is
symmetric relatively $c=$1, where the spectral width is maximum. The
maximum $\Delta$ lowers (rises) and shifts towards $c=$0 ($c=$2) for
the positive (negative) $b\rightarrow 0$ (Fig. \ref{fig10}). When
$b> -4.5$, the $+$ branch disappears for $a=$0 and $c=$1.

\begin{figure}
\includegraphics[width=8.5cm]{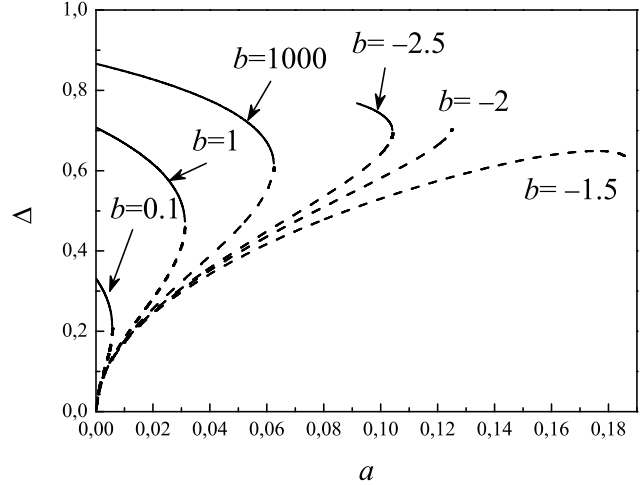}
\caption{\label{fig11} The CDS spectral half-widths for the $+$
(solid curves) and $-$ (dashed curves) branches in dependence on $a$
for the different $b$ and $c=$1. }
\end{figure}

Fig. \ref{fig11} demonstrates the dependencies of the spectral
half-width on $a$ for a varied $b$ and a fixed $c$. As a result of
larger energy and chirp, the $+$ branch (solid curves) has a larger
spectral width, which decreases with $a$ and the positive $b$
verging towards zero. The region of the $+$ branch existence
shortens with $b \rightarrow$0 (also, see Fig. \ref{fig1}). A
negative $b$ expands the $+$ branch region towards a larger $a$.
However, such a region is disconnected with $a=$0, if $b>-$4.5 for
$c=$1 (see the region for $b=-$2.5 in Fig. \ref{fig1}). The
existence of this minimum $c$ providing the CDS with $a=$0 is a
result of asymptotical behavior of the zero isogain in Fig.
\ref{fig5}. Physically, absence of the limit $a \rightarrow$0 can
mean that such a CDS is not able to develop from the vacuum of Eq.
(\ref{GL}). Fig. \ref{fig11} demonstrates that the $+$ branch
disappears completely, when $b \rightarrow -$2.

The $-$ branch has a lower spectral width, which increases with $a$
(Fig. \ref{fig11}, dashed curves). There exists some maximum $a$
(for a fixed $c$), where both branches merge.

\section{\label{compar} Numerical simulation of the CDS}

The above obtained approximate solution for the CDS has to be
verified numerically. With this purpose, a symmetrized split-step
Fourier method is used for numerical solving of Eq. (\ref{GL}). The
temporal greed contains 2$^{16}$ points, and the nonlinear
propagation is simulated in the time domain using a fourth-order
Runge-Kutta method. Total propagation distance consists of
$\geq$10$^4$ steps.

The simulations demonstrate, that the necessary factor providing the
CDS stability is a dependence of $\sigma$ on $E$. Such a dependence
is chosen to be in the form presented in Section \ref{soliton} (the
$\delta$ parameter equals to 0.5) \cite{k2}. Then, the $E$ parameter
in a master diagram can be easily replaced by the $E^*$ one, but the
difference between $E$ and $E^*$ is small and, therefore,
insignificant for further consideration.

The simulated spectra of the CDS are shown by open circles and
squares in Figs. \ref{fig8},\ref{fig9} for the parameters $a$, $b$
and $c$ corresponding to the points $\alpha$, $\beta$, $\gamma$ and
$\zeta$ in Figs. \ref{fig3},\ref{fig4},\ref{fig5}. The agreement
between the analytical and numerical results is perfect. Moreover,
the numerical results demonstrate that the CDS is really
three-parametric and its parameters scale in accordance with the
rules of Section \ref{soliton}. This means that the normalized
parameters and profiles of the CDSs are identical for the identical
sets of $(a, b, c)$. For instance, two parametric sets: i) $b=$20,
$a=$0.01, $\beta/\alpha=$30, $\zeta=0.1\gamma$, $E^*=$280 $\kappa
\sqrt{\kappa \zeta}/\gamma^2$, $\kappa=0.033 \gamma$ [e.g., a 100 nJ
Ti:sapphire oscillator with $\alpha=$2.5 fs$^2$ and $\gamma=$4.55
MW$^{-1}$ \cite{k2}]; and ii) $\beta/\alpha=$40, $E^*=$430 $\kappa
\sqrt{\kappa \zeta}/\gamma^2$, $\kappa=0.025 \gamma$ correspond to
the single point $\alpha$ in Fig. \ref{fig3}. This is the $-$
branch, and the analytical (solid curve) as well as numerical (gray
open circles) profiles coincide in Fig. \ref{fig9}. A single
difference between the numerical and analytical spectra is that the
former ones have gently smoothed edges. One has note, that a
scalability of the CDS resembles the property of a true soliton,
which is a solution with not fixed parameters \cite{a1}.

The numerically obtained stability borders are shown in Figs.
\ref{fig3},\ref{fig4},\ref{fig5} by open squares. The stability
condition is $\sigma>$0, that provides a vacuum stability. One can
see, that both analytical and numerical borders coincide.

It is of interest to compare the analytical results with the
numerical ones presented in \cite{a4,a5}. There is a difference
between the parametric sectors considered in \cite{a4,a5} and in
this work. The case of $\beta \rightarrow$0 lies beyond a validity
of the analytical model under consideration, which requires $\beta
\gg \alpha$. If $\beta$ approaches $\alpha$ and then tends to zero
(as well as if $\kappa$ prevails over $\gamma$), the spectrum edges
become smooth \cite{a5,k3} rather than truncated. The scaling rules
of Section \ref{soliton} and the requirement of $c<$2 can get broken
in this case \cite{m3}.

Nevertheless, i) strong scalability of $E$ with $\beta$, as well as
both ii) existence of maximum $\beta$ and iii) minimum $\kappa$
providing a stable soliton suggest that the solutions analyzed in
\cite{a4,a5} belong to the $+$ branch (here we consider only normal
dispersions). In conformity with \cite{a4}, the stable CDS exists
within the region of normal dispersion ($\beta>$0) for both positive
and negative $\chi$ (Figs. \ref{fig1},\ref{fig2}). A fast
disappearance of the CDS with the increase of $b<0$ \cite{a4} is the
characteristic feature of the $+$ branch (Fig. \ref{fig1}).

\begin{figure}
\includegraphics[width=8.5cm]{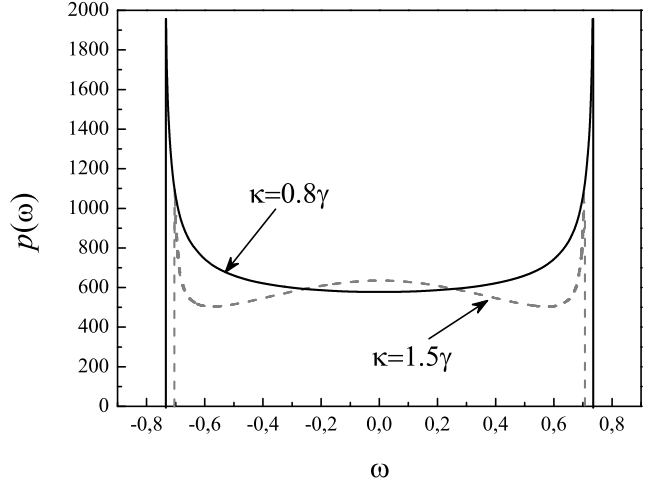}
\caption{\label{fig12}  The analytical CDS spectra for the $+$
branch at the points $\delta$ ($\kappa=$0.8$\gamma$; black solid
curve) and $\varepsilon$ ($\kappa=$1.5$\gamma$; gray dashed curve)
in Fig. \ref{fig4}. $b=$0.2, $\beta/\alpha=$6.25,
$\zeta=$0.002$\gamma$, and $\sigma=$0.1.}
\end{figure}

As expected, the CDS of \cite{a5} belongs to the $+$ branch (the
points $\delta$ and $\varepsilon$ in Fig. \ref{fig4}). The
corresponding analytical spectra are shown in Fig. \ref{fig12}
[$p(\omega)$ for the black solid curve is re-scaled for
convenience]. Both analytical spectra match with the numerical ones
in Fig. 3 of \cite{a5} with the exception of the smoothed edges for
the latter owing to $\kappa \geq \gamma$. Such a smoothing enhances
for the $-$ branch (the point $\epsilon '$ in Fig. \ref{fig4})
because a chirp is lower for this branch.

When $\kappa <
\gamma$ , the spectrum is concave (black solid curve in Fig.
\ref{fig12}) like that in Fig. \ref{fig8} for $b=$0.2. When $\kappa$
exceeds $\gamma$, the new type of a spectral shape appears: the
concave-convex one (gray dashed curve in Fig. \ref{fig12}). Such
spectra have been studied numerically in \cite{a5} for
$\kappa>\gamma$.

\section{Conclusion}

In conclusion, approximate chirped solitary pulse solutions of the
cubic-quintic nonlinear CGLE have been constructed analytically
under condition of domination of a dispersion over a spectral
dissipation. The solutions are three-parametric and easily traceable
within a whole parametric space, which has been represented in the
form of  the \textit{master diagrams}. The solutions are divided
into two branches, which differ in their energies and scaling
properties. It is found, that the chirped dissipative solitons under
consideration have truncated spectra with the concave, convex and
concave-convex tops. Numerical analysis and comparisons with the
existing results have demonstrated, that the approximate analytical
solutions are stable and highly-accurate. The obtained results are
of interest, in particular, for a development of both solid-state
and fiber laser oscillators aimed to a generation of femtosecond
pulses with over-microjoule energy.

\begin{acknowledgments}

Author thanks Boris Malomed for pointing out the solutions of
\cite{m1,m2}. This work was supported by the Austrian Fonds zur
F\"{o}rderung der wissenschaftlichen Forschung (FWF project P20293).
\end{acknowledgments}


\begin{thebibliography}{}
\bibitem{m3} perturbative analysis of the limit of $\alpha, \gamma, \chi \rightarrow
0$ is presented in \cite{m2}, where the soliton-like kink-antikink
solutions of Eq. (\ref{GL}) have been found.
\end{thebibliography}
\end{document}